\documentstyle[prb,aps,eqsecnum]{revtex}
\begin{document}

\def\A{{\bf A}}
\def\H{{\bf H}}
\def\x{{\bf x}}
\def\y{{\bf y}}
\def\k{{\bf k}}
\def\R{{\bf R}}
\def\r{{\bf r}}

\def\su{\uparrow}
\def\sd{\downarrow}

\draft
\title{Spin paramagnetism in d-wave superconductors}

\author{D. Tanaskovi\'c}
\address{Institute of Physics, P.O. Box 57, 11080 Belgrade, Yugoslavia}

\author{Z. Radovi\'c}
\address{Faculty of Physics, University of Belgrade, P.O. Box 368, 11001 Belgrade, Yugoslavia}

\author{L. Dobrosavljevi\'c-Gruji\'c}
\address{Institute of Physics, P.O. Box 57, 11080 Belgrade, Yugoslavia}

%\date{}
\maketitle

\begin{abstract}

The Ginzburg-Landau equations are derived from the microscopic theory
for clean layered superconductors with $d_{x^2-y^2}$ pairing symmetry,
including the Pauli paramagnetism effect. The upper critical
field $H_{c2}$ parallel to the $c$ axis is calculated. A comparison with the
experimental data for YBCO suggests that,  relative to the orbital effect,
the Pauli paramagnetism contribution to $H_{c2}$ is significant.
The reversible magnetization $M$  in high magnetic fields
  is also calculated, showing strong
temperature dependence of the slope $dM/dH$,
 as a consequence of the spin paramagnetism. A simple expression for the
  high temperature spin
 susceptibility is derived, in a good agreement with the Knight shift
  measurements on  YBCO.
\end{abstract}

\pacs{PACS: 74.72.-h; 74.25.Ha}

\section{Introduction}
Among the most important properties of the cuprate superconductors are their
 magnetic properties, reflecting the highly anisotropic layered structure and
  unconventional pairing mechanism and symmetry.\cite{Tinkham} Nowadays,
 the d-wave pairing symmetry in hole doped cuprates,\cite{Annet} and
quasi-two-dimensional (2D) nature of superconductivity,\cite{Harshman}
 are well established. Although the pairing mechanism is not known,
 the weak coupling BCS model for 2D d-wave superconductivity,\cite{MakiBCS}
 and the corresponding Ginzburg-Landau (GL) approximation,\cite{Ren}$^-$\cite{Shiraishi}
  are remarkably successful. 

The existence of
 superconducting phase in very
 high magnetic fields and layered structure of high-$T_c$ cuprates,
 makes the
 paramagnetic effect in the superconducting state much more important than
 in conventional superconductors.\cite{Yang,Won0}
The purpose of this paper is to derive GL equations for layered d-wave
 superconductors including the Pauli paramagnetism effect, giving simple
  analytical expressions for the upper critical field, magnetization,
   and spin susceptibility, suitable for comparison with experiments.

The GL equations for conventional (isotropic 3D s-wave) superconductors were
 first derived from  BCS theory by Gor'kov.\cite{Gor'kov} For 2D clean
  superconductors with d-wave pairing GL equations are derived
   by Ren {\it et al.},\cite{Ren} and extended by Won and Maki,\cite{Maki-Won}
    and   Shiraishi {\it et al.},\cite{Shiraishi} to include the higher order
     derivative terms. However, in the above references the magnetic field
    influence on the electron spins has been neglected. In their classical
 papers Maki and Tsuneto studied the effect of the
  Pauli paramagnetism in conventional superconductors.\cite{Maki-Tsuneto,Maki1}
   Their results mainly refer to the dirty limit.

In many underdoped and overdoped cuprates, where the upper critical fields
 $H_{c2}$ and transition temperatures $T_c$ are relatively low, the upward,
  positive curvature in the temperature dependence has been
   observed,\cite{Mackenzie,Osofsky} unlike the
 Werthamer-Helfand-Hohenberg   theory prediction for the conventional
  superconductors.\cite{Werthamer} A number of theoretical models for
   this unusual behavior has been proposed, involving, for example, the
 influence of the scattering by magnetic impurities and  of
  inhomogeneities,\cite{Kresin} the presence of  $d_{xy}$ pairing
   symmetry,\cite{Tachiki} and the higher Landau levels
  effect.\cite{Dukan}

In the optimally doped cuprates, the temperature dependence of $H_{c2}$
 parallel to the c axis appears to be  qualitatively the same as in the
 conventional superconductors. However, high values of  $H_{c2}$ and the
 irreversibility effects in optimally doped cuprates pose an obstacle to the
study of the field induced transition to the normal state.\cite{Blatter}
The relevant magnetization  measurements of $H_{c2}$ on high quality single
 crystals of $\rm{YBa_2Cu_3O_{7-\delta}}$ are performed by Welp
  {\it et al.}\cite{Welp} in a temperature interval of about 8 K below $T_c$.
   Recently, Nakagawa {\it et al.} \cite{Nakagawa,Nakagawa2} for field parallel
 and O'Brien {\it et al.} \cite{O'Brien} for field perpendicular to the $c$
  axis, reported data  from GHz transport measurements up to 150 T
 of $H-T$ phase diagram for $\rm{YBa_2Cu_3O_{7-\delta}}$
 thin films in the whole temperature range. 

Yang and Sondhi,\cite{Yang} and
 Won {\it et al.} \cite{Won0} studied theoretically the paramagnetic state of
 $d_{x^2-y^2}$ superconductors, neglecting the coupling of the magnetic
 field to the orbital motion of  electrons in the superconducting planes.
  For perpendicular field, measurements support their theory,
 strongly suggesting that $H_{c2}$ is limited by the spin paramagnetism below
       a certain characteristic temperature $T^*\sim 0.85T_c$.\cite{O'Brien}
 For parallel field the role of the spin paramagnetism should
 be clarified with regard to the orbital effect.

Magnetization measurements on YBCO in intermediate fields  of Sok {\it et al.},\cite{Sok}
are in a good agreement with the GL-like Hao and
 Clem model.\cite{Hao} In higher fields, where the magnetic phase diagram
 contains a vortex fluid,\cite{Blatter} one can presume that the
  diamagnetism of the vortex fluid resembles closely to that of an ideal Abrikosov mixed
  state at temperatures not too close to $T_c$.\cite{Hao1} In this regime,
  the measurements on cuprates should have the same slope in different fields,
   with strong temperature dependence due to the spin paramagnetism.

 Electronic spin susceptibility measurements provide evidence about
  the pairing state.\cite{Pines} Recent ESR Knight shift measurements on YBCO
   of J\' anossy {\it et al.}\cite{Janossy} clearly show the temperature
  dependence characteristic for 2D d-wave superconductivity, different from
  the BCS theory prediction for 3D s-wave superconductors.\cite{Yosida}

In Sec.~II we derive the GL equations for clean 2D  $d_{x^2-y^2}$
 superconductors in the magnetic field parallel to the $c$ axis, extending
  the procedure used by Ren {\it et al.} \cite{Ren} to include the effect of
 the spin paramagnetism. We calculate the upper critical field
  and discuss the importance of the spin paramagnetism regarding
  the experimental data for YBCO in Sec.~III. Within the model of layered
 superconductors as a stack of identical 2D conducting planes, we calculate
 the reversible magnetization in the Abrikosov approximation, Sec.~IV, and
  the spin susceptibility, Sec.~V. Section VI contains the conclusion.

\section{Ginzburg-Landau equations}
Including the interaction of the electron spins with the magnetic field, the four coupled Gor'kov equations are

\begin{equation}\label{2.1}
\left[ i \omega_n-\frac{1}{2m} (-i \nabla +\frac{e}{c} \A)^2-\mu_B H+\mu \right] G_{\su \su}(\x ,\x',\omega_n)-\int d\x''\Delta_{\su \sd}(\x,\x'')F_{\sd \su}^+(\x'',\x',\omega_n)=\delta(\x-\x'),
\end{equation}

\begin{equation}\label{2.2}
\left[ i \omega_n-\frac{1}{2m} (-i \nabla +\frac{e}{c} \A)^2+\mu_B H+\mu\right] G_{\sd \sd}(\x ,\x',\omega_n)-\int d\x''\Delta_{\sd \su}(\x,\x'')F_{\su \sd}^+(\x'',\x',\omega_n)=\delta(\x-\x'),
\end{equation}

\begin{equation}\label{2.3}
\left[ -i \omega_n-\frac{1}{2m} (i \nabla +\frac{e}{c} \A)^2+\mu_B H+\mu\right] F_{\sd \su}^+(\x ,\x',\omega_n)+\int d\x''\Delta_{\sd \su}^+(\x,\x'')G_{\su \su}(\x'',\x',\omega_n)=0,
\end{equation}

\begin{equation}\label{2.4}
\left[ -i \omega_n-\frac{1}{2m} (i \nabla +\frac{e}{c} \A)^2-\mu_B H+\mu\right] F_{\su \sd}^+(\x ,\x',\omega_n)+\int d\x''\Delta_{\su \sd}^+(\x,\x'')G_{\sd \sd}(\x'',\x',\omega_n)=0.
\end{equation}
Here, $G_{\alpha \alpha}(\x ,\x' ,\omega_n)$ and $F_{\alpha \beta }^+(\x, \x', \omega_n)$ are normal and anomalous Green's functions respectively,\cite{Abrikosov}  $\omega_n=(2n+1)\pi T$ are the Matsubara frequencies ($k_B=\hbar=1$), $\mu $ is the chemical potential and $\mu_B$ is the Bohr magneton. The magnetic field $\H =\mbox{curl} \A$ is perpendicular to the superconducting plane. The order parameter $\Delta_{\su \sd}^+$ is defined by self-consistency equation

\begin{equation}\label{2.5}
\Delta_{\su \sd}^+(\x,\x'')=V(\x-\x'')\langle \Psi_{\su}^{\dagger}(\x t) \Psi_{\sd}^{\dagger}(\x''t) \rangle=-V(\x-\x'')T\sum_{\omega_n} F_{\su \sd}^+(\x ,\x'',\omega_n),
\end{equation}
where $V(\x -\x'')$ is the effective pairing interaction. $\Delta_{\sd \su}^+(\x,\x''), \Delta_{\su \sd}(\x,\x'')$ and $\Delta_{\sd \su}(\x,\x'')$ are defined analogously.

From Eqs. $(\ref{2.1})-(\ref{2.5})$, to the third order in $\Delta$, we find

\begin{eqnarray}
\lefteqn{\Delta_{\su \sd}^+(\x,\y) =
V(\x-\y)T \sum_{\omega_n} \left\{ \int d\x' d\x'' \tilde G_{\su \su}(\x',\x,-\omega_n)\Delta_{\su \sd}^+(\x',\x'')
\left[ \tilde G_{\sd \sd}(\x'',\y,\omega_n) \right. \right. } \nonumber
\\
& & \left. \left. - \int d\x_1 d\x_2  \tilde G_{\sd \sd}(\x'',\x_1,\omega_n)  \Delta_{\sd \su}(\x_1,\x_2)\int d\x_3 d\x_4 \tilde G_{\su \su}(\x_3,\x_2,-\omega_n)\Delta_{\su \sd}^+(\x_3,\x_4) \tilde G_{\sd \sd}(\x_4,\y,\omega_n )\right] \right\} . \label{2.6}
\end{eqnarray}
Here $\tilde G_{\sigma}$ is the Green function in the normal state

\begin{equation}\label{2.7}
\tilde G_{\sigma}
(\x,\x',\omega_n)= \sum_{\k}{\frac{e ^{i\k (\x-\x')}}{i\omega_n -\sigma \mu_B H -\epsilon_{\k} }} e ^{-i \frac{e}{c} \A(\x)\cdot(\x-\x')},
\end{equation}
where $\sigma =\pm 1$ corresponds to spin up and down.
The classical approximation is used, taking account of the orbital effect by a change of phase, and $\epsilon_{\k}=(\k^2 /2m)-\mu$.

In the notation of Ref.~5, we find

\begin{eqnarray}
\lefteqn{\Delta^*(\R ,\k)=\sum_{\k'} V(\k-\k')T\sum_{\omega_n}\left[ \frac{1}{\omega_n^2+\epsilon_{\k'}^2}
+\frac{\epsilon_{\k'}^2-3\omega_n^2}{(\omega_n^2+\epsilon_{\k'}^2)^3}\mu_B^2 H^2
+ \frac {\epsilon_{\k'}^2-3\omega_n^2}{(\omega_n^2+\epsilon_{\k'}^2)^3} \frac{1}{4m^2}(k_x'^2\Pi_x^2+k_y'^2\Pi_y^2)\right. } \nonumber \\
& & \left. {}-\frac{1}{4m}\frac{\epsilon_{\k'}}{(\omega_n^2+\epsilon_{\k'}^2)^2}\Pi^2 \right] \Delta^*(\R,\k')-\sum_{\k'}V(\k-\k')T\sum_{\omega_n}\frac{1}{(\omega_n^2+\epsilon_{\k'}^2)^2}|
\Delta^*(\R,\k')|^2 \Delta^*(\R,\k'), \label{2.8}
\end{eqnarray}
where $\Delta_{\sd \su}^+ = \Delta_{\su \sd}^+\equiv \Delta^*,\, \R=(\x+\y)/2$ is the center of mass coordinate, and ${\bf \Pi}=-i\partial /\partial \R-2e\A/c $. The Fourier transform is performed with respect to the relative coordinate  $\r=\x-\y$, and $(\pi T)^2\gg (\mu_B H)^2$ is assumed.\cite{Maki-Tsuneto}

For $d_{x^2-y^2}$ pairing, we take
\begin{equation}\label{2.9}
V(\k-\k')=V_0(\hat k_x^2-\hat k_y^2)(\hat k_x'^2-\hat k_y'^2),
\end{equation}
and
\begin{equation}\label{2.10}
\Delta^*(\R,\k)=\Delta_0^*(\R)(\hat k_x^2-\hat k_y^2),
\end{equation}
where $\hat \k$ is the unit vector in the direction of $\k$.

From Eq. (\ref{2.8}), we obtain for $T$ near $T_c$ the first GL equation
\begin{equation}\label{2.11}
\Delta_0^*(\R)=\frac{1}{2} N(0)V_0\Delta_0^*\ln \frac{2C \omega_c}{\pi T}- \frac{7\zeta(3)}{8}\frac{1}{\pi^2 T_c^2}N(0)V_0\left[ \frac{1}{8} v_F^2\Pi^2\Delta_0^*+\mu_B^2 H^2 \Delta_0^* +\frac{3}{8}|\Delta_0^*|^2 \Delta_0^* \right] ,
\end{equation}
where $\omega_c$ is the energy cut-off, $C=1.78$ is the Euler constant, and 2D density of states for a given spin orientation $N(0)=m/2\pi$. This equation is the same as in Ref.~5, with an additional term containing the Bohr magneton $\mu_B$.

In the same approach, from the quantum mechanical equation for a single electron of spin $\hat{\bf s}$\cite{Landau}
\begin{equation}\label{2.121}
{\bf j}=-\frac{ie}{2m}\left[ (\nabla \Psi^*)\Psi-\Psi^*\nabla \Psi \right] -\frac{e^2}{mc}\A \Psi^* \Psi -2c\mu_B \mbox{curl}(\Psi^* \hat{\bf s}\Psi) ,
\end{equation}
the superconducting current density is

\begin{eqnarray}
\, {\bf j}_s(\x)=\frac{eT}{2im}\sum_{\omega_n}\sum_{\sigma}\int d\x_1d\x_2d\x_3d\x_4
\Delta(\x_1,\x_2)\Delta^*(\x_3,\x_4)\tilde G_{-\sigma}(\x_3,\x_2,-\omega_n)\nonumber \\
\,  {}\times (\nabla_{\x}-\nabla_{\y})\left. [\tilde G_{\sigma}(\x ,\x_1,\omega_n)
\tilde G_{\sigma}(\x_4,\y,\omega_n)]\right|_{\y=\x} \nonumber \\
\, {}+cT\mu_B\mbox{curl}\sum_{\omega_n}\sum_{\sigma=-1}^1{\vec \sigma}\int d\x_1d\x_2d\x_3d\x_4\Delta(\x_1,\x_2)\Delta^*(\x_3,\x_4)  \nonumber \\
\, {}\times \tilde G_{\sigma}(\x ,\x_1,\omega_n)
\tilde G_{-\sigma}(\x_3,\x_2,-\omega_n)\tilde G_{\sigma}(\x_4,\x,\omega_n), \label{2.122}
\end{eqnarray}
where $\vec \sigma$ is the unit vector orthogonal to the superconducting plane, $\vec \sigma=\sigma \H/H$. The first term in Eq. (\ref{2.122}) is the diamagnetic current ${\bf j}_s^d$, and the second term is due to the difference between the paramagnetic current associated with the electron spins in the normal and the superconducting state, ${\bf j}_s^{spin}$. Note that the total current is ${\bf j}={\bf j}_s+2N(0)\mu_B^2\mbox{curl}\H$, where the second term is the paramagnetic current in the normal state. Introducing the center of mass, performing the Fourier transform of $\Delta $ with respect to the relative coordinates, and after integrations similar as in Ref.~5, we get
\begin{eqnarray}
\, {\bf j}_s=\frac{eT}{m^2}\sum_{\omega_n}\int \frac{d\k}{(2\pi)^2}\Delta^*(\R,\k)\frac{1}{\omega_n^2+\epsilon_{\k}^2}\left\{  \left. \frac{m\k}{i\omega_n-\epsilon_{\k}}[\nabla_{\k'} \cdot {\bf \Pi}^*\Delta(\R,\k')]\right|_{\k'=-\k} \right. \nonumber\\
\left. {}-\frac{2\k}{(i\omega_n-\epsilon_{\k})^2}[\k \cdot {\bf \Pi}^*\Delta(\R,-\k)]\right\} +\mbox{H.c.} \nonumber \\
{}+cT\mu_B\mbox{curl}\sum_{\omega_n} \int\frac{d\k}{(2\pi)^2}\Delta(\R,-\k)\Delta^*(\R,\k) 2\mu_B\H \frac{\epsilon_{\k}^2-3\omega_n^2}{(\omega_n^2+\epsilon_{\k}^2)^3}. \label{2.124}
\end{eqnarray}
Here, in ${\bf j}_s^d$ the spin paramagnetic effect is neglected, as well as the orbital effect in ${\bf j}_s^{spin}$.\cite{Maki1}

For $d_{x^2-y^2}$ pairing, Eq. (\ref{2.10}), after integration over $\k$, for $T$ near $T_c$, and assuming $(\pi T)^2\gg (\mu_BH)^2$, the second GL equation for the total supercurrent density is \cite{error}

\begin{equation}\label{2.12}
{\bf j}_s=\frac{7\zeta(3)N(0)mv_F^2}{32\pi^2 T_c^2}\left[ -\frac{ie}{m}\left( \Delta_0 \frac {\partial \Delta_0^*}{\partial \R}-\Delta_0^*\frac{\partial \Delta_0}{\partial \R}\right) -\frac{4e^2}{mc}\A |\Delta_0^*|^2 -\frac{8c}{mv_F^2}\mu_B^2 \mbox{curl} (\H |\Delta_0^*|^2 )\right] .
\end{equation}

Finally, for a stack of identical 2D conducting planes in the magnetic field parallel to the $c$ axis, the corresponding GL equations in the standard form are
\begin{equation}\label{2.13}
\alpha \psi + \beta |\psi |^2\psi+\frac{1}{2m}{{\bf \Pi}^*}^2\psi+\eta H^2\psi=0,
\end{equation}

\begin{equation}\label{2.14}
 {\bf{j}}_s=-\frac{i e}{m}(\psi \frac{\partial\psi^*}{\partial{\bf R}}-\psi^*\frac{\partial \psi}{\partial{\bf R}})-\frac{4e^2}{mc}\A |\psi|^2-2c\eta \mbox{curl} (\H |\psi |^2),
\end{equation}
where
\begin{equation}\label{2.15}
\alpha=\frac{16\pi^2 T_c^2}{7\zeta(3)mv_F^2}\frac{T-T_c}{T_c},
\end{equation}
\begin{equation}\label{2.16}
\beta=\frac{48\pi^2T_c^2}{7\zeta(3)N^*(0)m^2v_F^4},
\end{equation}
and
\begin{equation}\label{2.17}
\eta =\frac{4}{mv_F^2}\mu_B^2.
\end{equation}
Here $N^*(0)=N(0)/\delta$, $\delta $ being the average spacing between the planes.\cite{Harshman}

In this model, the GL equations for a d-wave, clean and layered superconductor are of the same form as in the isotropic and clean 3D s-wave case. However, besides the factor $2/3$ in parameters $\alpha $, $\beta $, $\eta $, the quantities $m,v_F,N(0)$ refer to the effective mass, Fermi velocity and density of states in 2D conducting $ab$ planes. Therefore, the coherence length is $\xi=\xi_{ab}=(1/2m|\alpha |)^{1/2}$, the penetration depth $\lambda=\lambda_{ab}=(mc^2\beta /16\pi e^2|\alpha |)^{1/2}$, and the GL parameter
\begin{equation}\label{2.18}
\kappa=\frac{\sqrt 6 cT_c}{ev_F^2}\sqrt{\frac{\pi }{7\zeta(3)N^*(0)}}.
\end{equation}
The free energy density corresponding to Eqs. (\ref{2.13}) and (\ref{2.14}) is

\begin{equation}\label{2.19}
F=F_n+\alpha|\psi|^2+\frac{\beta}{2}|\psi|^4+\frac{1}{2m}|{\bf \Pi}^*\psi|^2+\frac{H^2}{8\pi}+\eta H^2|\psi|^2.
\end{equation}

\section{Upper critical field}

Near the second order phase transition to the normal phase, from the linearized Eq. (\ref{2.13})
\begin{equation}\label{3.11}
\alpha \psi + \frac{1}{2m}{{\bf \Pi}^*}^2\psi+\eta H^2\psi=0,
\end{equation}
we obtain for the upper critical magnetic field parallel to the $c$ axis

\begin{equation}\label{3.1}
H_{c2}=\frac{ev_F^2}{8c\mu_B^2}\left[ \sqrt{1+\frac{256\pi^2c^2\mu_B^2T_c^2}{7\zeta(3)e^2v_F^4}(1-\frac{T}{T_c})}-1 \right] .
\end{equation}
The slope at $T_c$ is
\begin{equation}\label{3.3}
\left. \frac{dH_{c2}}{dT}\right| _{T_c}= -\frac{16\pi^2cT_c}{7\zeta(3)ev_F^2},
\end{equation}
and the GL expression without the Pauli paramagnetism correction is simply
\begin{equation}\label{3.31}
H_{c2}^0=T_c\, |dH_{c2}/dT|_{Tc}\, (1- \frac{T}{T_c}) .
\end{equation}

We illustrate our results using the experimental data for YBCO.\cite{Welp,Nakagawa}
 Samples are in the clean limit, and for fields above 1T the spin-orbit scattering can
  be neglected.\cite{Yang} Taking the slope $dH_{c2}/dT|_{Tc}=-1.9$ T/K from the
 magnetization measurements on optimally doped YBCO ($T_c=92$ K) of
  Welp {\it et al.},\cite{Welp} we find that at $T= 0.7\, T_c$ the spin
   pair breaking lowers the critical field for 10\%, Fig. 1(a).
However, it is evident that a larger slope corresponds to the data of
 Nakagawa {\it et al.}\cite{Nakagawa} We obtain with Eq. (\ref{3.1}) a 
  good fit of the experimental data for  $T\gtrsim 0.5\, T_c$,
   taking the slope
  $dH_{c2}/dT|_{Tc}=-2.6$ T/K and $T_c=84.3$ K, Fig. 1(b). For this slope the paramagnetic
  correction is $-15$\% at $T=0.7\, T_c$.

   At low temperatures, where GL theory is not
  applicable, $H_{c2}(T)$ becomes saturated.
  From the data of Nakagawa {\it et al.},\cite{Nakagawa}
  $H_{c2}(0)\approx 110$ T. The Won and Maki expression $H_{c2}(0)=-0.63\, T_c\,  dH_{c2}/dT|_{Tc}$,\cite{Maki-Won} relevant for 2D d-wave
 superconductors, for $T_c=84.3$ K and the slope
$-2.6$ T/K gives $H_{c2}(0)=138$ T. 
 Therefore, the spin paramagnetism, not taken into account in Ref.~7, but relevant in this case because the Clogston limit is 155 T,  
should lower $H_{c2}(0)$ for $20$\%.
 However, note that $H_{c2}$ measurements strongly depend on
  experimental techniques.\cite{Nakagawa2,O'Brien}

The Fermi velocities corresponding to $dH_{c2}/dT|_{Tc}=-1.9$ T/K,
 $T_c=92$ K and $dH_{c2}/dT|_{Tc}=-2.6$ T/K, $T_c=84.3$ K are
  $v_F=10.1 \cdot 10^6$ cm/s and $v_F=8.3 \cdot 10^6$ cm/s, respectively.
 The latter value is closer to that obtained from independently measured
 Fermi energy 
  and the effective mass,\cite{Harshman}
  giving $v_F=(7.6\pm 0.9)\cdot 10^6$ cm/s.

\section{Reversible magnetization}

The GL Eqs. (\ref{2.13}), (\ref{2.14}) in the dimensionless form are

\begin{equation}\label{4.1}
(i\frac{\nabla}{\kappa}- {\bf a})^2f=f-|f|^2f-\gamma h^2 f ,
\end{equation}
and

\begin{equation}\label{4.2}
-\mbox{curlcurl}\, {\bf a} =|f|^2{\bf a} -\frac{i}{2\kappa}(f^*\nabla f-f\nabla f^*)+\gamma \mbox{curl}({\bf h} |f|^2) ,
\end{equation}
where

\begin{equation}\label{4.3}
\gamma=\frac{32\pi N^*(0)\mu_B^2}{3}(1-\frac{T}{T_c}).
\end{equation}
The dimensionless quantities are introduced with $\psi=\sqrt{-\alpha /\beta} f, \, \R=\lambda \tilde{\R}, \, 2e\xi\A /c={\bf a}$, and $2e\xi \lambda \H /c={\bf h}$.\cite{Sarma} Eqs. (\ref{4.1})-(\ref{4.3}) are of the same form as for isotropic 3D s-wave superconductor,\cite{Maki-Tsuneto} with different $\kappa $, and $\gamma $, Eqs. (\ref{2.18}) and  (\ref{4.3}).

Following the Abrikosov approach,\cite{Sarma} in the vicinity of the upper critical field parallel to the $c$ axis of the layered superconductor, we obtain

\begin{equation}\label{4.4}
h=h_e-\frac{|f|^2}{2\kappa'},
\end{equation}

\begin{equation}\label{4.41}
\overline{|f|^2}=\frac{1}{\kappa'}\frac{h_{c2}-h_e}{(1-1/2\kappa'^2)\beta},
\end{equation}

\begin{equation}\label{4.5}
b=h_e-(h_{c2}-h_e)\frac{1}{(2\kappa'^2-1)\beta}.
\end{equation}
Here, $h_e$ is the external magnetic field, $\beta=\overline{|f|^4}/\overline{|f|^2}^2$, the induction $b$ is the mean value of the microscopic  magnetic field $h$, and

\begin{equation}\label{4.6}
\kappa'=\frac{\kappa}{\sqrt{1+4\kappa^2\gamma}}.
\end{equation}
The dimensionless expression for the upper critical field, Eq. (\ref{3.1}), is
\begin{equation}\label{4.61}
h_{c2}=\frac{\sqrt{1+4\kappa^2\gamma}-1}{2\gamma \kappa}.
\end{equation}
We emphasize that in Eq. (\ref{4.5}) magnetization due to the spin polarization of the normal electrons is not included. As in 3D s-wave case,\cite{Maki-Tsuneto} the effect of the Pauli paramagnetism on $b$ is included simply by the scaling of $\kappa \rightarrow \kappa'$. Therefore, the vortex lattice is again equilateral triangular lattice with $\beta=1.16$. The temperature dependence of $\kappa '$ leads to the characteristic variation of the magnetization, $4\pi m=b-h_e$, with temperature. From Eqs. (\ref{4.5})-(\ref{4.61}), for $\kappa \gg 1$, the slope in physical units
\begin{equation}\label{4.62}
\frac{dM}{dH_e}=\frac{1+4\kappa^2 \gamma}{4\pi \beta (2\kappa^2-1)}
=\left. \frac{dM}{dH_e}\right|_{Tc}
\left[ 1+\frac{256\pi^2c^2\mu_B^2T_c^2}{7\zeta(3)e^2v_F^4}(1-\frac{T}{T_c}) \right]
\end{equation}
is linearly decreasing with temperature, instead of the temperature
 independent slope in the GL approach without paramagnetic correction.
  Due to higher $T_c$ and smaller $v_F$, this effect is not negligible
 in high-$T_c$ cuprates, in contrast to the conventional superconductors.
  For $dH_{c2}/dT|_{Tc}=-1.9$ T/K, we obtain a strong variation $(dM/dH_e)/(dM/dH_e)|_{Tc}=1+1.4(1-T/T_c)$, or $1+2.6(1-T/T_c)$ for $dH_{c2}/dT|_{Tc}=-2.6$ T/K. 

In high-$T_c$ superconductors the influence of the vortex lattice melting,
 and the strong fluctuation effects in the vicinity of $T_c$ also affect
  $dM/dH_e$ temperature dependence.\cite{Blatter} However, because of the strong
   paramagnetic effect, the characteristic
    increase of $dM/dH_e$ with decreasing temperature, predicted by
     Eq. (\ref{4.62}), should be dominant for $H\gtrsim 0.5\, H_{c2}$, and  $0.5\lesssim T/T_c\lesssim 0.9$. This effect may be compensated only by the strong temperature dependence of the effective parameter $\kappa_2$ within the extended GL theory approach.\cite{Eilenberger}

\section{Spin susceptibility}

The current density due to the spin paramagnetism is
\begin{equation}\label{4.7}
{\bf j}_s^{spin}(\r)=c\, \mbox{curl}\; ({\bf M}_s^{spin}(\r)-{\bf M}_n(\r)),
\end{equation}
where ${\bf M}_s^{spin}(\r)$ and ${\bf M}_n(\r)=2N^*(0)\mu_B^2{\bf H}$ are the magnetization due to the spin polarization in the superconducting and in the normal state.\cite{Maki1}
From Eq. (\ref{4.7}) and the last term in Eq. (\ref{2.14}),

\begin{equation}\label{4.8}
M_s^{spin}=M_n-\frac{8\mu_B^2}{mv_F^2}H|\psi|^2.
\end{equation}

In the weak magnetic field $|\psi|^2=-\alpha /\beta $, and the spin susceptibility $\chi_s =M_s^{spin}/H$ is

\begin{equation}\label{4.9}
\chi_s=\chi_n\left[ 1-\frac{4}{3}(1-\frac{T}{T_c})\right] .
\end{equation}
This simple GL expression for the spin susceptibility of the layered d-wave superconductor is in a very good agreement with the microscopic theory result of Won and Maki,\cite{MakiBCS} in a large temperature range  $T\gtrsim 0.6\, T_c$. Analogously, the susceptibility in the GL approach for an isotropic 3D s-wave, as well as for 2D s-wave, superconductor is

\begin{equation}\label{4.10}
\chi_s=\chi_n\left[ 1-2(1-\frac{T}{T_c})\right] ,
\end{equation}
in the agreement with the Yosida result from BCS theory.\cite{Yosida} Therefore, d-wave and s-wave symmetries of the order parameter lead to different slopes of the spin susceptibility curve in the high temperature regime.

This is illustrated in Fig. 2, in comparison with the ESR Knight shift data for YBCO of J\' anossy {\it et al.}\cite{Janossy} Small discrepancy of the experimental data and the theoretical predictions at high temperatures could be the consequence of the strong coupling effects,\cite{Janossy} or of the presence of a small s-wave component of the order parameter which is possible in YBCO.\cite{Maki-Won}

\section{Concluding remarks}

The GL equations for layered d-wave clean superconductor are derived including
 the Pauli paramagnetism effect. The parallel upper critical field and the
  reversible magnetization in the Abrikosov approximation are calculated.
   The results are of the same form as obtained by Maki and Tsuneto in
    isotropic 3D s-wave case, with changed $\kappa $ and $\gamma $.
  The expression for $H_{c2}$ gives an useful correction to
 the standard GL result, providing the correct determination
  of $dH_{c2}/dT|_{Tc}$ for high-$T_c$ superconductors from the fit of experimental data in
  the relatively large range of temperatures below $T_c$. From the comparison
   with the experimental data of for
   $\rm{YBa_2Cu_3O_{7-\delta}}$ thin films,\cite{Nakagawa} we obtain $dH_{c2}/dT|_{Tc}=-2.6$ T/K, 
    and conclude that the effect of Pauli paramagnetism on $H_{c2}$ parallel to
    the $c$ axis is significant in comparison to the orbital effect, $-15$\% at
     $T=0.7\, T_c$. At zero temperature, this effect should be greater, about $-20$\%.  The strong temperature dependence of the
      magnetization slope $dM/dH_e$ due to the spin paramagnetism influence, should be experimentally detectable in high
      field measurements.
We have also derived the GL expression for  the Knight shift, reflecting the pairing symmetry in accordance with the experimental data.

\begin{figure}
    \caption{ }
Temperature dependence of the parallel upper critical field of YBCO.
Solid curve: GL result with the paramagnetic correction, Eq. (\ref{3.1}).
The weak coupling microscopic theory result without the paramagnetic
correction (dotted curve),\cite{Maki-Won} and the corresponding GL result,
Eq. (\ref{3.31}), (dashed line) are shown for comparison. Solid circles
represent $H_{c2}$ measurements of Welp {\it et al.}\cite{Welp} (a), and
Nakagawa {\it et al.}\cite{Nakagawa} (b).
\end{figure}

\begin{figure}
    \caption{ }
Temperature dependence of the normalized spin susceptibility. Solid
lines: GL approximation, Eqs. (\ref{4.9}) and (\ref{4.10}). Dotted
curves: the weak coupling microscopic theory
calculations.\cite{MakiBCS,Yosida} Top curves correspond to 2D d-wave,
and bottom curves to 3D s-wave case. Solid circles: data of J\' anossy
{\it et al.}\cite{Janossy} for YBCO.
\end{figure}


\begin{thebibliography}{22}
\bibitem{Tinkham} M. Tinkham, {\it Introduction to Superconductivity} (McGraw Hill, New York, 1996), ch. 9.
\bibitem{Annet} J. F. Annet, N. Goldenfeld, and A. J. Legget in: D. M. Ginsberg (Ed.), {\it Physical Properties of High Temperature Superconductors}, v. 5, (World Scientific, Singapure, 1996).
\bibitem{Harshman} D. R. Harshman and A. P. Mills, Jr., Phys. Rev. B {\bf 45}, 10684 (1992).
\bibitem{MakiBCS} H. Won and K. Maki, Phys. Rev. B {\bf 49}, 1397 (1994).
\bibitem{Ren} Y. Ren, J. H. Xu, and C. S. Ting, Phys. Rev. Lett. {\bf 74}, 3680 (1995).
\bibitem{Ren1} J. H. Xu, Y. Ren, and C. S. Ting, Phys. Rev. B {\bf 52}, 7663 (1995).
\bibitem{Maki-Won} H. Won and K. Maki, Phys. Rev. B {\bf 53}, 5927 (1996).
\bibitem{Shiraishi} J. Shiraishi, M. Kohmoto, and K. Maki, Phys. Rev. B {\bf 59}, 4497 (1999).
\bibitem{Yang} K. Yang and S. L. Sondhi, Phys. Rev. B {\bf 57}, 8566 (1998).
\bibitem{Won0} H. Won, H. Jang, and K. Maki, cond-mat/9901252.
\bibitem{Gor'kov} L. P. Gor'kov, Zh. Eksp. Teor. Fiz. {\bf 36}, 1918 (1959) [Sov. Phys. JETP {\bf 9}, 1364 (1960)].
\bibitem{Maki-Tsuneto} K. Maki and T. Tsuneto, Progr. Theoret. Phys. {\bf 31}, 945 (1964).
\bibitem{Maki1} K. Maki, Phys. Rev. {\bf 148}, 362 (1966).
\bibitem{Mackenzie} A. P. Mackenzie, S. R. Julian, G. G. Lonzarich, A. Carrington, S. D. Hughes, R. S. Liu, and D. C. Sinclair, Phys. Rev. Lett. {\bf 71}, 1238 (1993).
\bibitem{Osofsky} M. S. Osofsky, R. J. Soulen, Jr., S. A. Wolf, J. M. Broto, H. Rakoto, J. C. Ousset, G. Coffe, S. Askenazy, P. Pari, I. Bozovic, J. N. Eckstein, and  G. F. Virshup, Phys. Rev. Lett. {\bf 71}, 2315 (1993).
\bibitem{Werthamer} N. R. Werthamer, E. Helfand, and P. C. Hohenberg, Phys. Rev. {\bf 147}, 295 (1966).
\bibitem{Kresin} Yu. N. Ovchinnikov and V. Z. Kresin, Phys. Rev. B {\bf 54}, 1251 (1996);  Europhys. Lett., {\bf 46}, 794 (1999).
\bibitem{Tachiki} T. Koyama and M. Tachiki, Physica C {\bf 263}, 25 (1996).
\bibitem{Dukan} M. Rasolt and Z. Te\v sanovi\'c, Rev. Mod. Phys. {\bf 64}, 709 (1992); S. Dukan and O. Vafek, Physica C {\bf 309}, 295 (1998).
\bibitem{Blatter} G. Blatter, M. V. Feigel'man, V. B. Geshkenbein, A. I. Larkin, and V. M. Vinokur, Rev. Mod. Phys. {\bf 66}, 1125 (1994).
\bibitem{Welp} U. Welp, W. K. Kwok, G. W. Crabtree, K. G. Vandervoort, and J. Z. Liu,  Phys. Rev. Lett. {\bf 62}, 1908 (1989).
\bibitem{Nakagawa} H. Nakagawa, T. Takamasu, N. Miura, and Y. Enomoto, Physica B {\bf 246-247}, 429 (1998).
\bibitem{Nakagawa2} H. Nakagawa, N. Miura, and Y. Enomoto, J. Phys.: Condens. Matter {\bf 10}, 11571 (1998).
\bibitem{O'Brien} J. L. O'Brien, H. Nakagawa, A. S. Dzurak, R. G. Clark, B. E. Kane, N. E. Lumpkin, N. Miura, E. E. Mitchell, J. D. Goettee, J. S. Brooks, D. G. Rickel, and R. P. Starrett,  cond-mat/9901341.
\bibitem{Sok} J. Sok, M. Xu, W. Chen, B. J. Suh, J. Gohng, D. K. Finnemore, M. J. Kramer, L. A. Schwartzkopf, and B. Dabrowski, Phys. Rev. B {\bf 51}, 6035 (1995).
\bibitem{Hao} Z. Hao and J. R. Clem, Phys. Rev. Lett. {\bf 67}, 2371 (1991).
\bibitem{Hao1} Z. Hao, J. R. Clem, M. W. McElfresh, L. Civale, A. P. Malozemoff, and F. Holtzberg, Phys. Rev. B {\bf 43}, 2844 (1991).
\bibitem{Pines} D. Pines and P. Wr\' obel, Phys. Rev. B {\bf 53}, 5915 (1996).
\bibitem{Janossy} A. J\' anossy, T. Feh\' er, G. Oszl\' anyi, and G. V. M. Williams,  Phys. Rev. Lett. {\bf 79}, 2726 (1997).
\bibitem{Yosida} K. Yosida, Phys. Rev. {\bf 110}, 769 (1958).
\bibitem{Abrikosov} A. A. Abrikosov, L. P Gor'kov, and I. E Dzyaloshinski {\it Methods of Quantum Field Theory in Statistical Physics} (Dover, New York, 1975).
\bibitem{Landau} L. D. Landau and E. M. Lifshitz, {\it Quantum Mechanics} (Addison-Wesley,  Reading, 1958).
\bibitem{error} Note that a factor 2 is missing in Eq. (17), Ref. 5, and in Eqs. (4.3)-(4.4),  in Ref. 6.
\bibitem{Sarma} D. Saint-James, G. Sarma and E. J. Thomas, {\it Type II  Superconductivity} (Pergamon Press, Oxford, 1969).
\bibitem{Eilenberger} G. Eilenberger, Phys. Rev. {\bf 153}, 584 (1967); K. Maki and T. Tsuzuki, Phys. Rev. {\bf 139}, A868 (1965).

\end{thebibliography}
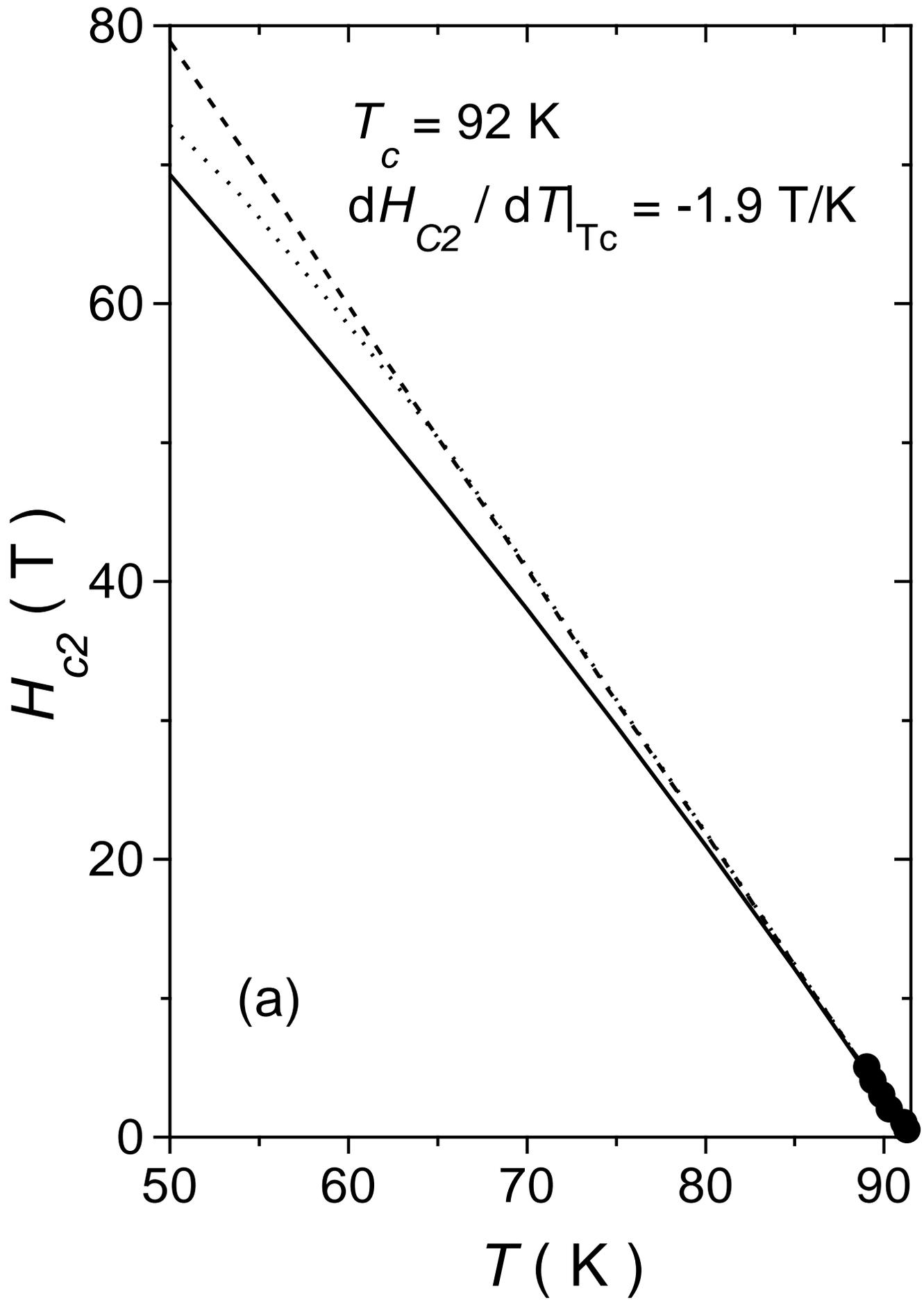
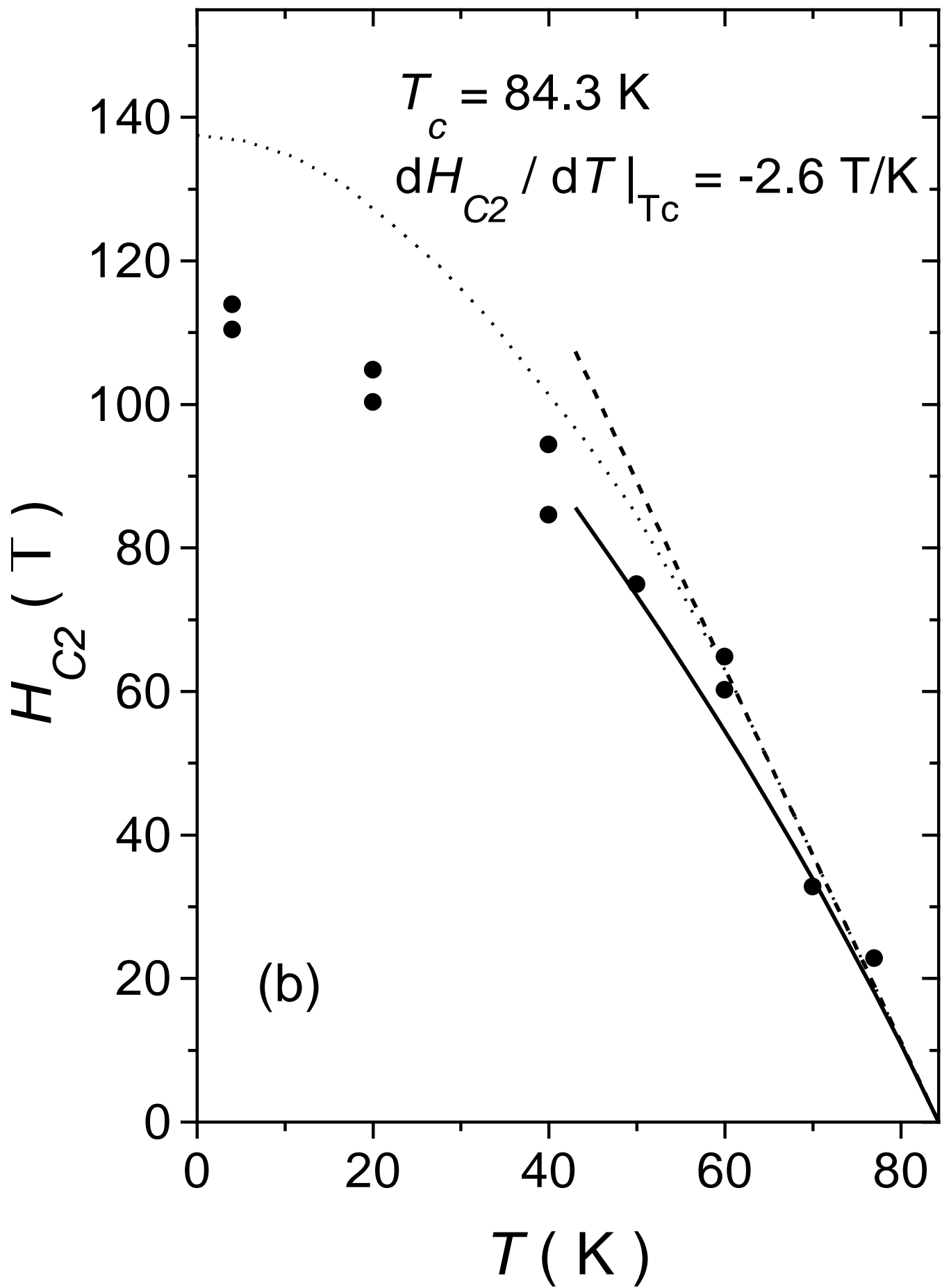
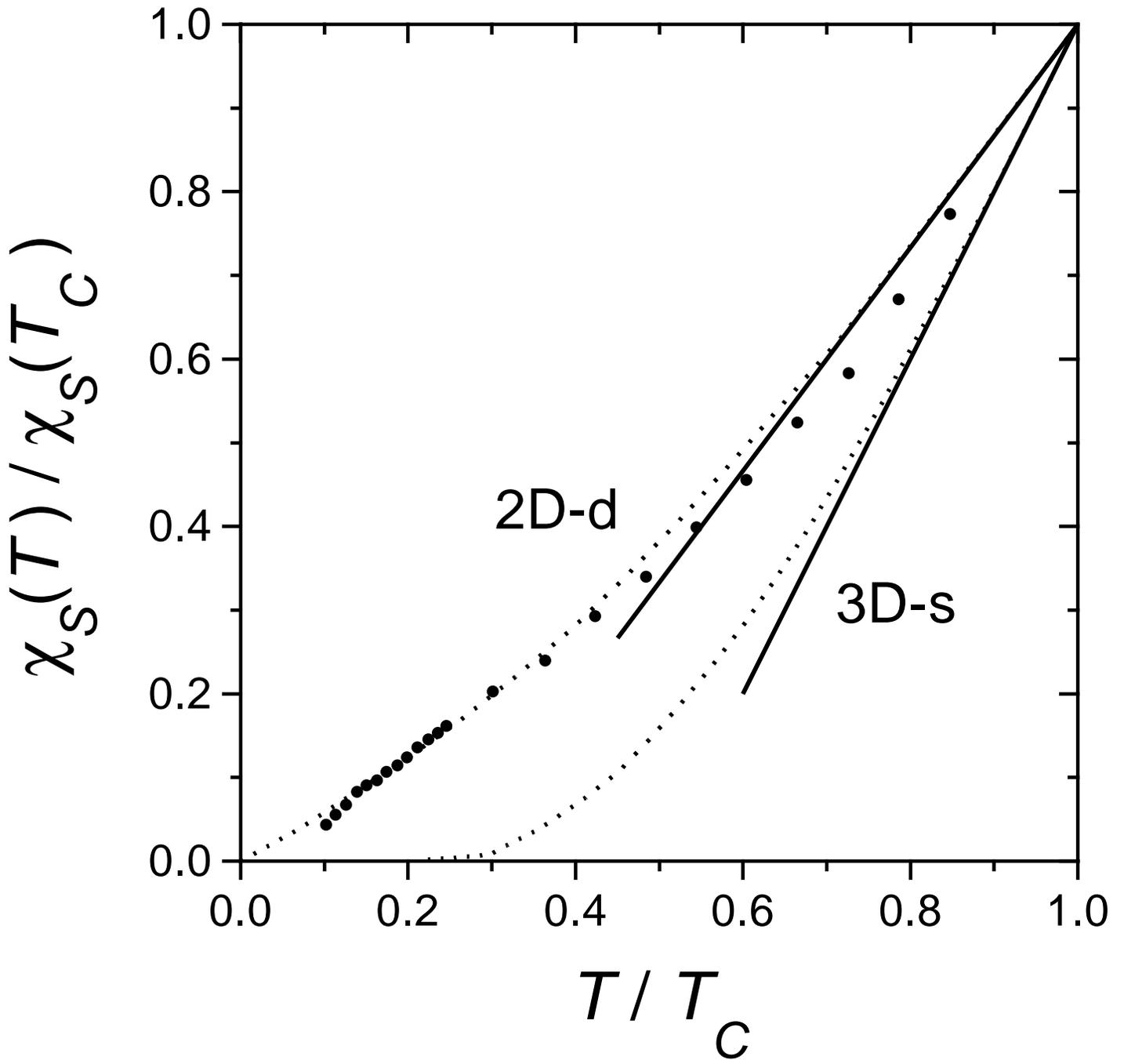
\end{document}